# Influenza Hospitalisations in England during the 2022/23 Season: do different data sources drive divergence in modelled waves? A comparison of surveillance and administrative data.

Jonathon Mellor[1], Rachel Christie[1], James Guilder[1], Robert S Paton[1], Suzanne Elgohari[1], Conall Watson[1], Sarah Deeny[1], Thomas Ward[1*]

1. UK Health Security Agency, Data Analytics and Science, Noble House, London

*Corresponding Author: Tom.Ward@UKHSA.gov.uk

## Abstract

### Background

Accurate and representative data is vital for precisely reporting the impact of influenza in healthcare systems. Northern hemisphere winter 2022/23 experienced the most substantial influenza wave since the COVID-19 pandemic began in 2020. Simultaneously, new data streams become available within health services because of the pandemic. Comparing these data, surveillance and administrative, supports the accurate monitoring of population level disease trends.

### Methods

We analysed admissions rates per capita from four different collection mechanisms covering National Health Service hospital Trusts in England over the winter 2022/23 wave. We adjust for difference in reporting and extracted key epidemic characteristics including the maximum admission rate, peak timing, cumulative season admissions and growth rates by fitting generalised additive models at national and regional levels.

### Results
By modelling the admission rates per capita across surveillance and administrative data systems we show that different data measuring the epidemic produce different estimates of key quantities. Nationally and in most regions the data correspond well for the maximum admission rate, date of peak and growth rate, however, in subnational analysis discrepancies in estimates arose, particularly for the cumulative admission rate.

### Interpretation

This research shows that the choice of data used to measure seasonal influenza epidemics can influence analysis substantially at sub-national levels. For the admission rate per capita there is comparability in the sentinel surveillance approach (which has other important functions), rapid situational reports, operational databases and time lagged administrative data giving assurance in their combined value. Utilising multiple sources of data aids understanding of the impact of seasonal influenza epidemics in the population.

## Introduction

Surveillance of influenza is crucial to monitor epidemiological trends and assess current and future pressures on primary and secondary health care services. A key aspect of influenza surveillance is the monitoring of hospitalisation rates, which require timely estimations of the number of patients admitted to hospital and the population those admitted draw from. Monitoring these rates is important as they give a measure of season severity, indicators of pressure that allow healthcare systems to adjust care delivery and the ability to compare the disease dynamics through time and across locations. Influenza hospitalisations in England are measured and monitored in several ways across different administrative data sources and active surveillance systems. These data are collected for a range of purposes, some are explicitly for influenza surveillance such as SARI Watch [1] or for example managing hospital payments, such as the Secondary Use Services (SUS) database [2].

Influenza is a transmissible respiratory virus that causes seasonal epidemics worldwide, leading to a high morbidity [3, 4] and mortality [5] as well as pressures on primary and secondary care services [6, 7]. The influenza virus has 4 types, only two of which, influenza A and B, tend to cause seasonal epidemics [8]. The most common symptoms of influenza are fever, body aches, fatigue, dry cough, sore throat, headaches [9] and the disease can cause life threatening complications, with greatest risk for those over 65 and under 2 years of age or with chronic health conditions [10, 11, 12]. Since the onset of the COVID-19 pandemic influenza transmission was limited by the impacts of non-pharmaceutical-interventions (NPI) [13, 14, 15], changes to contact networks [16], and increase in the size of the susceptible population [17], making a resurgent influenza epidemic wave of key importance for healthcare operational demands. The joint healthcare pressure of an influenza wave and a COVID-19 wave at the same time, co-circulating [18], posed an at the time unseen scenario for care providers in 2022, with a considerable risk of overstretched intensive care unit capacities [19].

In this research we compare four data sources of influenza hospitalisations over the winter 2022/23 season in England and explore the different characterisations of the epidemic wave they produce by modelling each data source. We produce estimates of key metrics including the size, timing, and rate of change of the influenza admissions rates at national and National Health Service (NHS) regional geographies.

## Methodology

This study focuses on the comparison of different data and surveillance sources for influenza admission rates, below we outline the different data and their key distinctions in measurement and purpose. Furthermore, in this study we compare different key epidemic qualities of interest across data sources. These metrics include the timing and magnitude of the wave peak, point of maximum growth rate and timing as well as the cumulative season admission rate. This is achieved by modelling the different data sources to adjust for differences in their temporal granularity, sample size and extract the key metrics from these models.

**Data**

**Secondary Uses Service – All Patient Care**

The secondary uses services data set is created from information relating to payment for activity undertaken whilst a patient is admitted to hospital. The administrative data is published monthly for the financial year so far and allows for additional data to be added to pre-existing patient records [20]. There are five data sets within Secondary Uses Services (SUS) database: Accident and Emergency (ECDS), admitted patient care (APC), Adult Critical Care, Outpatients and Maternity. The focus for this paper is the SUS APC data, which is made available with a 2-3 month reporting lag. The data is on a per consultant episode per trust basis and includes geographical information about the treatment site, as well as patient demographic information (such as age and patient residence). The data includes admission/discharge date, diagnosis, and procedure codes, so it is possible to identify the reasons a person is admitted and what procedures took places during their stay. This data covers all Integrated Care Boards in England [2, 21]. Influenza patients are identified using ICD-10 codes J09, J10 and J11 [22] as either primary or secondary diagnosis. Patients with COVID-19 emergency code (U071), or those individuals who tested positive for COVID-19 between fourteen days before, and one day after admission were classified as COVID-19 patients and excluded from this analysis. The COVID-19 tests were linked from the UKHSA Second Generation Surveillance Service (SGSS) [23]. The APC data is not necessarily always complete, reports are filed after discharge and episodes statistics may be backfilled, and diagnosis codes may not be recorded, so the quality of the data will vary across trusts [24]. Data was obtained on the 28 August 2023. Patients who receive an influenza diagnosis should have tested positive for the disease based on guidance [25], though without direct data linkage between patient and test this is challenging to verify. This data source tells us granular information about the patient; however, it tells us little about the disease itself, with no typing or sub-typing information recorded.

**SARI Watch**

UKHSA SARI Watch is a surveillance system designed to monitor respiratory disease hospitalisations. It includes weekly data for test confirmed influenza, RSV and COVID-19, and differentiates between hospital admissions to all levels of care and admissions to critical care (intensive care unit/high-dependency units ICU/HDU) [26] and provides the patient age range. The main difference between SARI Watch and SUS APC reporting is that SARI Watch is based on reporting within a week of admission for near real time surveillance whereas APC is upon completion of an admission. There are also important differences in the availability of subtyping information which is collected in SARI Watch to inform severity assessment and vaccine policy decision. There are also difference in definitions of influenza patients, in part this due to testing. SARI Watch contains patients which test positive for these diseases either through a point of care or laboratory test, as well as having clinical symptoms [26]. SUS APC definition is determined by clinical coding, which may have an associated test. SARI Watch provides additional information, including the influenza A subtype and type detected, which can support understanding of the current influenza virus landscape [26]. The SARI-Watch Sentinel admissions data is collected from a subset of trusts in England forming a sentinel network. In comparison to the passive collection of other data sources through administrative collection, the SARI data relies on recruitment of participating trusts and engagement to maintain the high quality of collection. The SARI

Watch Sentinel influenza admissions data we explore in this study contains a week identifier based on the ISO week system, trust identifier, as well as admission counts for influenza stratified by age band and influenza sub-type or type. Sentinel Trusts are re-recruited annually, with some joining and leaving each year, though there is inter-year consistency and sites were initially recruited using stratified sampling to ensure a representative sample of the whole of England [26]. This data is used as the accepted estimate of admissions rates as the UK Official Statistics, with a history of utility for monitoring respiratory illness. The near real time epidemiological data is also used for public health responsiveness throughout the season.

**National Health Service England Urgent and Emergency Care Situational Report**

National Health Service England (NHSE) urgency and emergency care data is provided by individual NHS trusts who deliver a daily situation report (SitRep) on urgent and emergency care (UEC) by 11am each day covering the previous 24 hours [27]. The administrative reporting process for many trusts is automated and is completed via web form [28]. Information on both hospital bed occupancy and patients in the past 24 hours who test positive for influenza and made available weekly. For comparison, we assume positive test patients in the last 24 hours are analogous to admissions. The occupancy counts are based on the latest snapshot in the previous 24 hours on the day of reporting (~8am) [29], which are divided into two categories: critical care (ICU/HDU beds) as well as general and acute. This system was created as a wider situation report for more than influenza; however, we have restricted this analysis to only this virus. The data collection of influenza was started in 2021 and so does not contain historic influenza pre-COVID-19 comparison benchmarks. The data source provides aggregate patient counts, rather than information of the disease itself, such as typing or sub-typing produced by tests, or relevant patient characteristics like age.

**Second Generation Surveillance System**

The Second Generation Surveillance System (SGSS) is an administrative database storing infectious disease test results in the UK [23]. Within the UK influenza is a notifiable causative agent according to the Health Protection (Notification) regulations (2010) [30]. Both positive and negative tests are required to be reported within seven days to SGSS by laboratories in 2022 [31], we are particularly interested in those reported by hospitals in this work. , including those of interest for this research within hospitals. The SGSS data was obtained on 12 September 2023. Within SGSS there are identifiers for the category of test (such as a rapid test or PCR), the influenza type (A or B, where the test allows this), the setting of the test (such as primary care, inpatient or emergency care), which organisation requested the test and some patient characteristics. We selected only tests requested by NHS Trusts, and within an inpatient setting as well as deduplicating the tests to be one per patient episode to get the date a patient first tested positive for influenza, giving a proxy for an influenza hospitalisation. This approach relies on complete records where the NHS Trust and setting can be identified, which is not always the case. The completeness of fields that may be used to link tests and accuracy of these identifiers varies by NHS Trust and over time. The SGSS data provides some information on the disease itself, such as influenza A or B and on the patient characteristics such as age.

| Data Source | Label | Admission Definition |
|---|---|---|
| Secondary Use Services – Admitted Patient Care | APC | Patient admitted to hospital with a primary or secondary diagnosis code of influenza. Excluding patients that have a primary or secondary diagnosis of COVID-19 or positive COVID-19 test within 14 days of admission. |
| NHS England Urgent and Emergency Care Situational Report | UEC | New inpatients in the past 24 hours with a laboratory confirmed positive influenza test. This will include both admissions due to influenza, and patients with influenza. |
| SARI Watch Sentinel | SARI | Patient admitted to hospital which meets clinical symptom criterion and has a confirmed positive influenza test. |
| Second Generation Surveillance System | SGSS | First positive test for an individual while an inpatient within a hospital. |

*Table 1. Definition of an influenza admission across APC, UEC, SARI and SGSS data. Admissions are considered as patients admitted into the acute hospital in either a general ward or critical care. Both PCR and molecular point of care tests are included.*

**Population Catchment Estimates**

To determine admission rates per capita, allowing us to compare aggregate estimates of admissions across disparate hospital groups, we need a population denominator. NHS Trusts do not have a clearly defined population they service, with multiple providers placed within administrative regions. Which hospital a patient attends emergency care in is dependent on location and choice [32]. This health seeking behaviour changes across age groups [33] and will vary for different infectious diseases.

To calculate the effective population of an NHS Trust we produced a proportionate mapping between NHS Trust and lower-tier local authorities (LTLA) utilising the same structure from the Office of Health Improvement and Disparities NHS Acute (Hospital) Trust Catchment Population experimental statistics [34]. We queried the APC database for all patients and then aggregated by trust and LTLA to produce the proportionate mapping.

With this mapping, we used the ONS 2019 local authority population estimates to produce a weighted sum of populations across trusts of their feeder LTLAs, giving an effective population catchment size for each hospital trust.

**Processing**

To compare across different datasets, were transformed into a consistent structure. Each dataset is transformed into a count of influenza admissions per day per NHS Trust that reported to the respective system. The population estimate for each NHS Trust is joined onto the conformed data. We define the influenza season as 02 October 2022 to 21 May 2023, epidemiological weeks 40 to 20.

The SARI Watch data are weekly counts; therefore, we convert to days by dividing by 7. As the collection is Monday-Sunday, the date is selected as the midpoint, Thursday. The NHS Trusts included are acute secondary care providers defined in the Estates Returns Information Collection data [35], with specialist Trusts removed. As the different sources are a mix of opt-in surveillance (SARI Watch), administrative secondary purpose collections (SUS, SGSS) and a newly set up reporting system (UEC), there are varying levels of Trust participation. We define an inclusion criteria for Trusts to ensure a minimum level of quality to reduce bias in the estimate. Trusts are excluded from the analysis using the following criteria:

1. The Trusts reported missing values over the whole study period (aggregate reports count: UEC, SARI).
2. The Trust reported a value of 0 for more than 90% of dates in the study period (aggregate reports: UEC SARI). The Trust did not report a test or admission for over 90% of dates in the study period (individual reports: APC, SGSS).

Criteria 1 is used to remove non-participating Trusts in aggregate reporting and criteria 2 is used to remove incorrect null returns, which bias downward the admission rate or Trusts that did not provide identifying details in individual details, such as an organisation code. The choice in threshold value for criteria 2 is shown in Supplementary Figure 1.

**Model**

To infer characteristics of the epidemic wave we need to quantify uncertainty due to sample sizes and adjust for reporting effects. Three of the four data sources within the analysis are daily, which introduces a day-of-week effect, with hospital admissions and/or tests dropping over weekends. We estimated the admissions per capita as a smooth function of time ($s(t)$) using a generalised additive model (GAM). The GAM used thin-plate splines through time with a negative binomial error structure, log-link function and model offset of the population size. A random effect $f_2$ was used to adjust for the day-of-week effects dow) in reporting for APC, UEC and SGSS data sources, but this was not needed for the weekly SARI data. The thin-plate spline $f_1$ is fit over time points $t$. The national models are defined in Equation 1 and 2, with the regional models defined by Equation 3 and 4, where i denotes the region, producing an independent spline fit for each region, with a complete pooled day-of-week adjustment. The population size is given as $population\ size(t)$ which is the sum of Trust catchment populations that reported admissions at $t$.

$$s(t) = log(admissions(t)) =$$
$$\beta_0 + f_1(t) + f_2(dow) + log\,(population\ size(t)) \quad \textit{Equation 1}$$

$$s(t) = log(admissions(t)) =$$
$$\beta_0 + f_1(t) + log\,(population\ size(t)) \quad \textit{Equation 2}$$

$$s(t) = log\,(admissions(t))$$
$$\beta_0 + f_1(t,i) + f_2(dow) + log\,(population\ size(t,i)) \quad \textit{Equation 3}$$

$$s(t) = log(admissions(t)) =$$
$$\beta_0 + f_1(t,i) + log\,(population\ size(t,i)) \quad \textit{Equation 4}$$

The models are fit using the R package *mgcv* [36]. From the models we extracted samples of the fit to quantify both model and data uncertainty as well as the first derivative of the epidemic wave, the growth rate, using a central finite difference approach with the R package *gratia* [37]. By taking the estimated smooth function of time $s(t)$, its derivative the daily growth rate $\frac{d\,s(t)}{dt}$ we calculate the instantaneous doubling time as $\frac{\log(2)}{\frac{d\,s(t)}{dt}}$.

**Comparison Metrics**

Day of week effects in admissions are confounders of the true underlying epidemic growth rate. As these are modelled as random effects in equations 1 and 2, we omit these weekly cycles in admission numbers from our presented per capita admission rates and growth rates. From the fit models we generate 1000 posterior samples for the admissions per capita and growth rate via an approximate multivariate normal method, using the *gratia* package [37]. From these posterior samples we calculate summary statistics across the draws from the model fit and use quantiles to capture the uncertainty. For each summary metric of the epidemic wave by data source, we produced a median estimate and 95% confidence interval.

To characterize the epidemic wave we produce estimates of key metrics of interest for the size, timing and rate of change of the influenza admissions. The maximum admission rate is the maximum value of each set of posterior draws across the wave. The peak date is the date at which the maximum admission rate occurs across the posterior draws. The cumulative admission rate is the sum of the daily admission rate over the wave for each set of posterior draws. We also perform inference on qualities of the epidemic wave first derivative, extracting the maximum growth rate, timings of change points and the length of peak, defined as the difference between the maximum and minimum growth rate.

## Results

**Reporting Coverage**

Working with an admission rate per capita and changing Trust participation means the population denominator can change over time for some datasets as participation evolves. The national count of Trusts reporting over the season are shown in Table 1, along with raw counts of the peak and cumulative admissions to demonstrate the scale of data source unadjusted for reporting variation, regional breakdowns are given in Supplementary Table 1. How the proportion of a population covered by the dataset changed through time is shown in Figure 1. APC consistently reports the highest coverage, at near 100% population catchment reporting, though this is not in real-time, with a three-month reporting lag. In England, the next highest is the UEC source which covers approximately 75% of the country. This proportion changes with time, with more Trusts starting to report part way through the season after the peak had passed driven by the Midlands and London regions. Regionally SGSS has reporting from 50% population coverage and above, though there is again regional heterogeneity. This proportion is driven by missing information in the database, causing exclusions due to criteria 2 – Trusts not reporting tests over 90% of the season. The SARI dataset has lowest coverage which is expected from a sentinel surveillance approach aiming for a representative sample of Trusts, rather than full coverage.

| Dataset | Trusts reporting (minimum – maximum) | Peak Admissions | Cumulative Admissions |
|---|---|---|---|
| APC | 115 - 115 | 1505 | 46043 |
| SARI | 12 - 19 | 227 | 1236 |
| SGSS | 78 - 78 | 1088 | 29606 |
| UEC | 89 - 105 | 1028 | 32706 |

Table 1. Unmodelled national summary of reported data across the different data sources. Counts are from the processed data after exclusion criteria are applied. Trust counts are the lowest and highest number of participating Trusts for a given report post exclusion criteria. Peak admissions are taken as the maximum admissions in each report and cumulative admissions the sum of all admissions reported. These metrics are not corrected for time varying participation and population catchment sizes. Regional breakdowns are given in Supplementary Table 1.

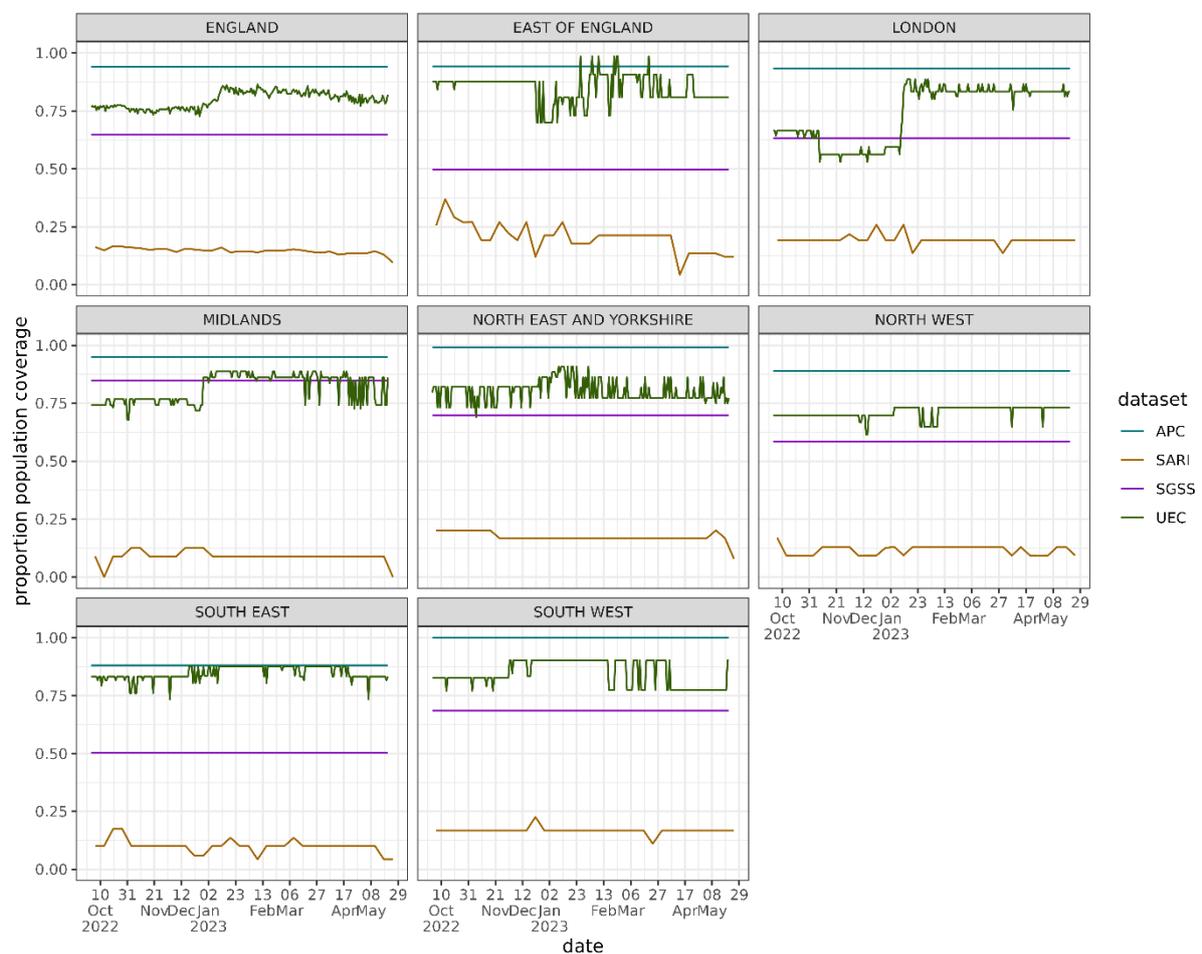

Figure 1. The population catchment of trusts reporting to each data source over time as a proportion of the total population in that geography. As individual level data sources, APC and SGSS do not change over time and are lower than 1.0 due to the inclusion criteria for reporting.

**Modelled Admissions and Doubling Time**

The winter 2022/23 seasonal influenza wave was early in the season with a fast growth and similarly fast decline. The modelled national admissions wave timing appears to correspond well between datasets shown in Figure 2A, though the magnitudes of the rates differ. There is high uncertainty in the admission rate for UEC, with SARI having consistently higher values outside of the peak wave period. There is a similar picture of agreement in the doubling times for the different datasets in Figure 2B. Each source is growing at a similar rate and had a fastest doubling time of approximately seven days. The unmodelled data with clear day-of-week effects and stochastic noise can be found in Supplementary Figure 2.

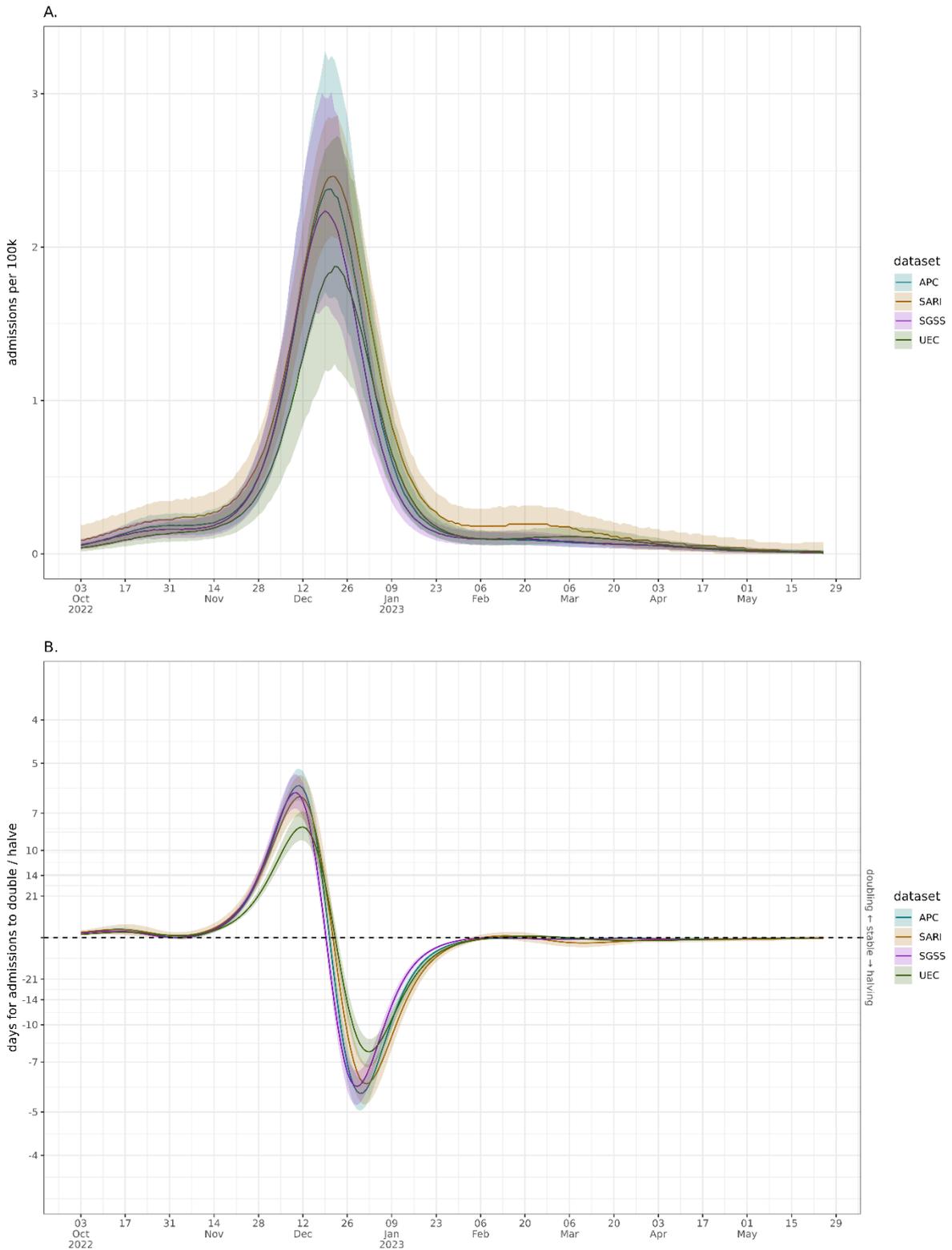

*Figure 2. National modelled admission per capita wave with weekly effect correction (sub-plot A) and the national admissions growth rate expressed as a doubling/halving time (sub-plot B) across the different data sources. Solid lines represent median model estimate and ribbons the 95% confidence interval.*

The estimated regional per capita rates in Figure 3 broadly show agreement in trend across the data sources. The exception to this is in London, where SARI estimates a notably higher rate with a gap between confidence intervals for much of the peak. UEC and SGSS estimated admission rates are substantially below the other two sources and appear to be lower than the other regions. The APC estimate in London is between these two different trends, more in line with other regions. While it is important not to over interpret growth rates at low counts (the early and late season) there is regional uncertainty across datasets in Figure 5 in the beginning and end of the season. This high uncertainty in the growth rate means it takes longer into the season to determine when a growth phase has started which would be when the confidence interval no longer crosses a doubling time of zero. As expected from the smaller sample sized approach there is more uncertainty in the SARI growth rate, though the direction agrees with other data.

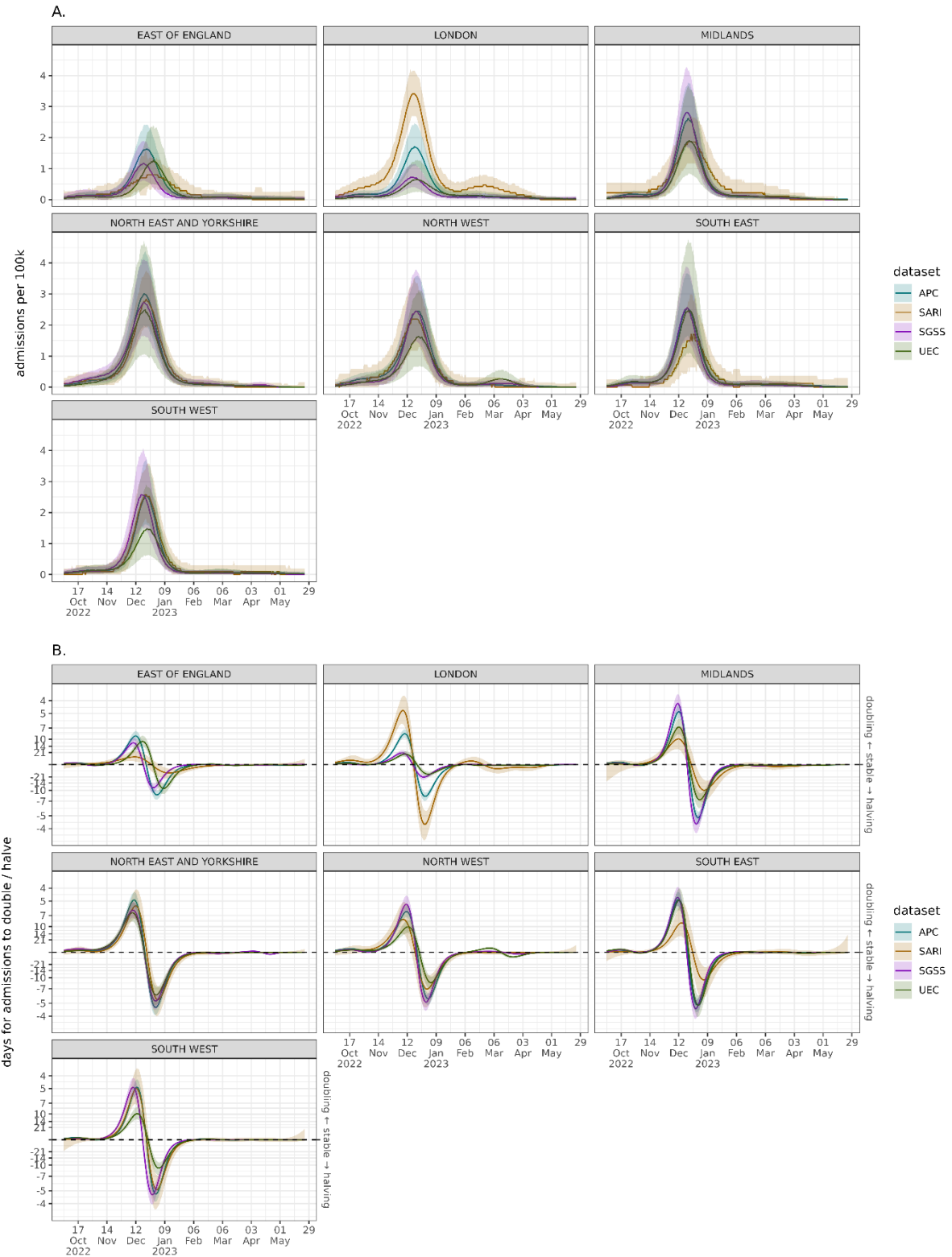

*Figure 3. Regional modelled admission per capita wave with weekly effect correction (sub-plot A) and the regional admissions growth rate expressed as a doubling/halving time (sub-plot B) across the different data sources. Solid lines represent median model estimate and ribbons the 95% confidence interval.*

## Epidemic Wave Peak

The peak admissions rate of an influenza wave is a key characteristic when comparing seasons, but as shown in Figure 4 there is variation in the estimates for the peak admissions rate regionally. Nationally, there is strong agreement on the maximum rate, with central estimates all between 2.4 and 3 admissions per 100k with overlapping confidence intervals. There is high uncertainty in the estimate produced by UEC across most regions, and as with other analysis we can see a disparity in datasets for estimating the peak in London. Excluding London and the East of England there is agreement in the magnitude of the peak which implies there may be different ascertainment rates or reporting in London and East of England.

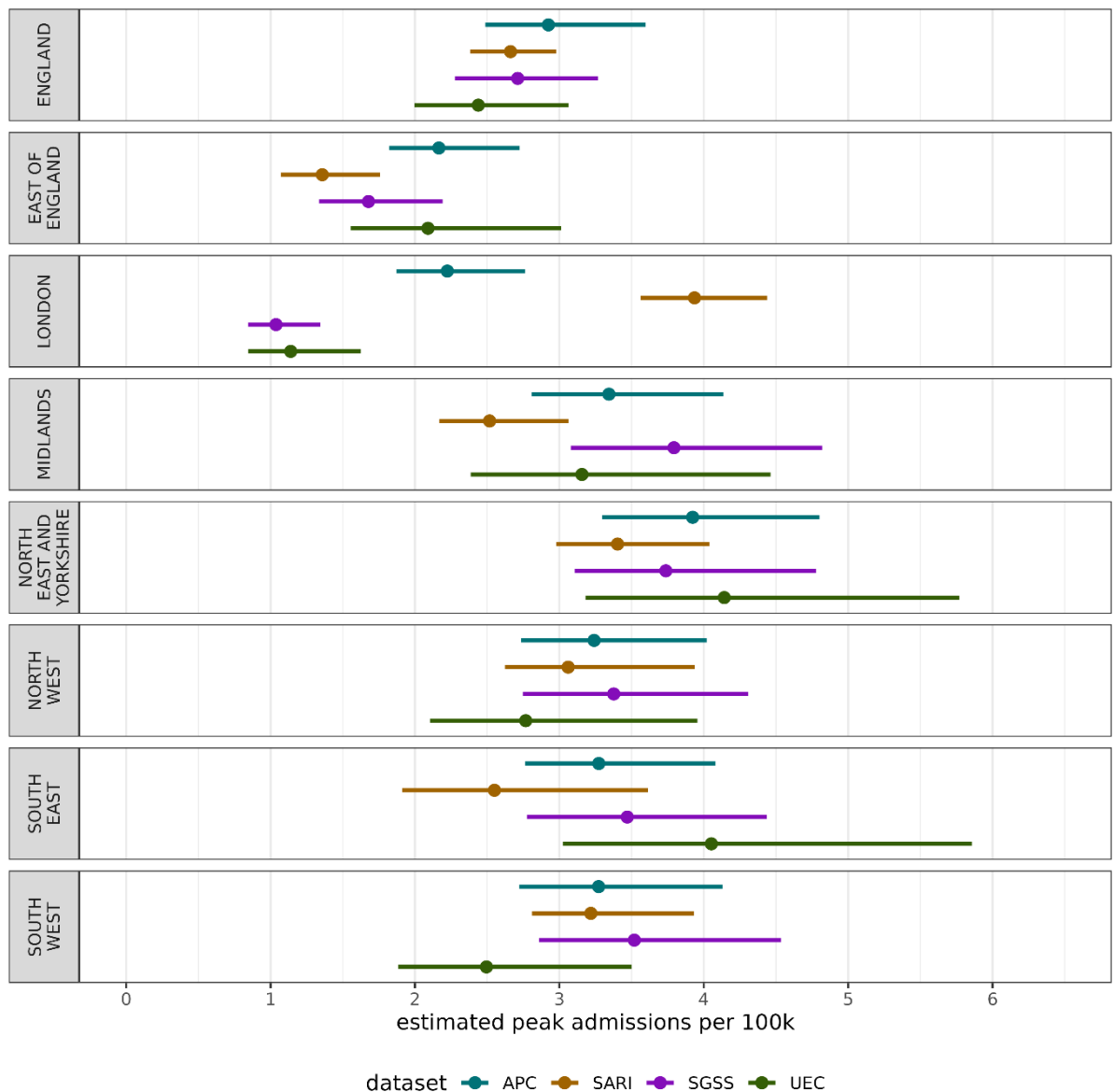

*Figure 4. The estimated maximum number of admissions per capita nationally and regionally across each data source at the peak of the epidemic wave for the winter 2022/23 season. The central point represents the median estimate, and the lines the 95% confidence interval.*

The timing of an epidemic peak is crucial for healthcare providers and public health officials to know. The analysis in Figure 5 shows that the different data sources provide similar estimates of the day of peak, centred around 22 December 2022. The lack of spatial variation in peak timing implies all regions experienced a similar epidemic which did not spread across the country, but rather occurred simultaneously. Due to the day-of-week adjustment (APC, SGSS, UEC) or weekly data (SARI) there is large uncertainty about when the peak occurred - with 95% intervals between 10-14 days nationally or larger regionally.

**Season Cumulative Admissions Rate**

While the timing and magnitude of the influenza admissions peak is a measure of system pressure, the burden on the healthcare system can be characterised over the whole season as the cumulative season admission rates explored in Figure 5. There is high variation in cumulative admission rate across regions and datasets. The clearest difference is the low total admissions in UEC and SGSS within London, and substantially higher rate in SARI. This London difference drives much of the national difference in cumulative rates.

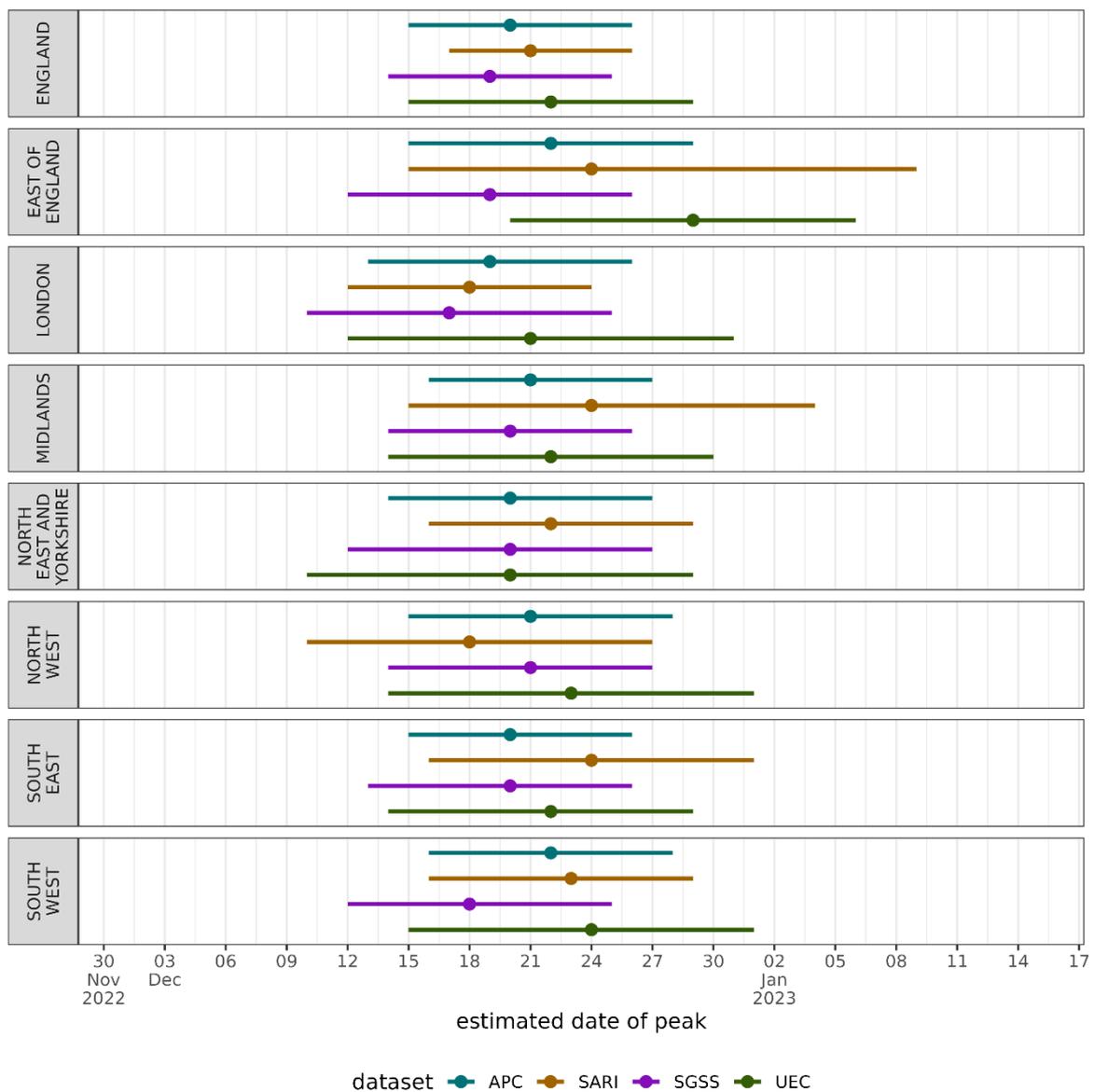

*Figure 5. The estimated date of the epidemic wave peaking nationally and regionally across each data source in the winter 2022/23 influenza season. The central point represents the median estimate, and the lines the 95% confidence interval.*

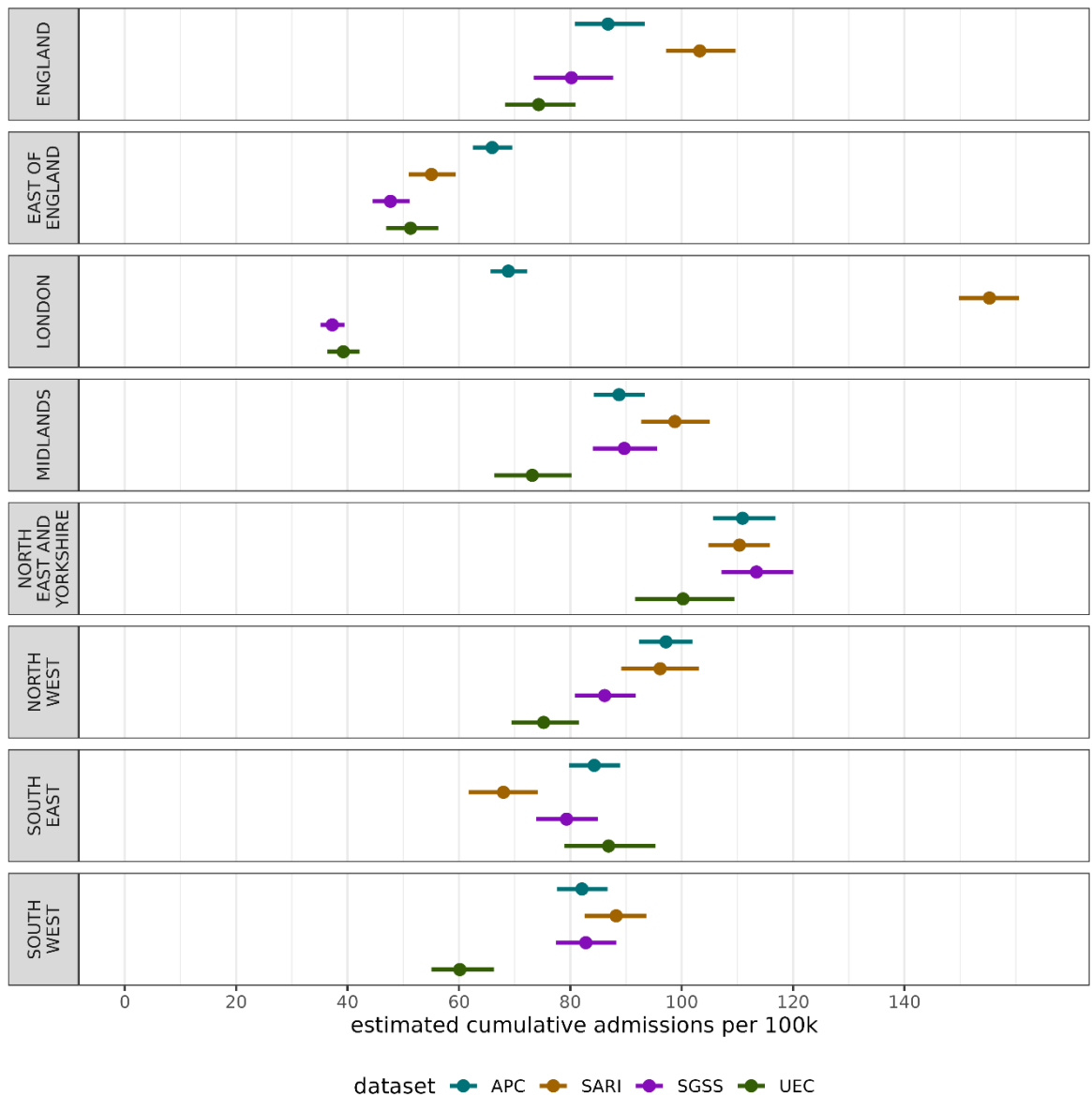

*Figure 6. The estimated cumulative admissions per capita nationally and regionally across each data source over the 2022/23 winter season in England. The central point represents the median estimate, and the lines the 95% confidence interval.*

**Speed of Doubling Time**

How quickly an epidemic is growing is an important metric for understanding it's progression – to understand the winter 2022/23 influenza wave's speed we extracted the maximum doubling time of the wave presented in Figure 7. Nationally, the epidemic wave was shown to double very quickly, with all datasets giving a maximum doubling time of near seven days. Regionally, there is much variation, with some datasets giving even shorter doubling times and some substantially longer, such as SGSS and UEC in London, and SARI in East of England. The estimated dates of fastest doubling and halving times are given in Supplementary Figure 3 and the estimated length of peaks in Supplementary Figure 4.

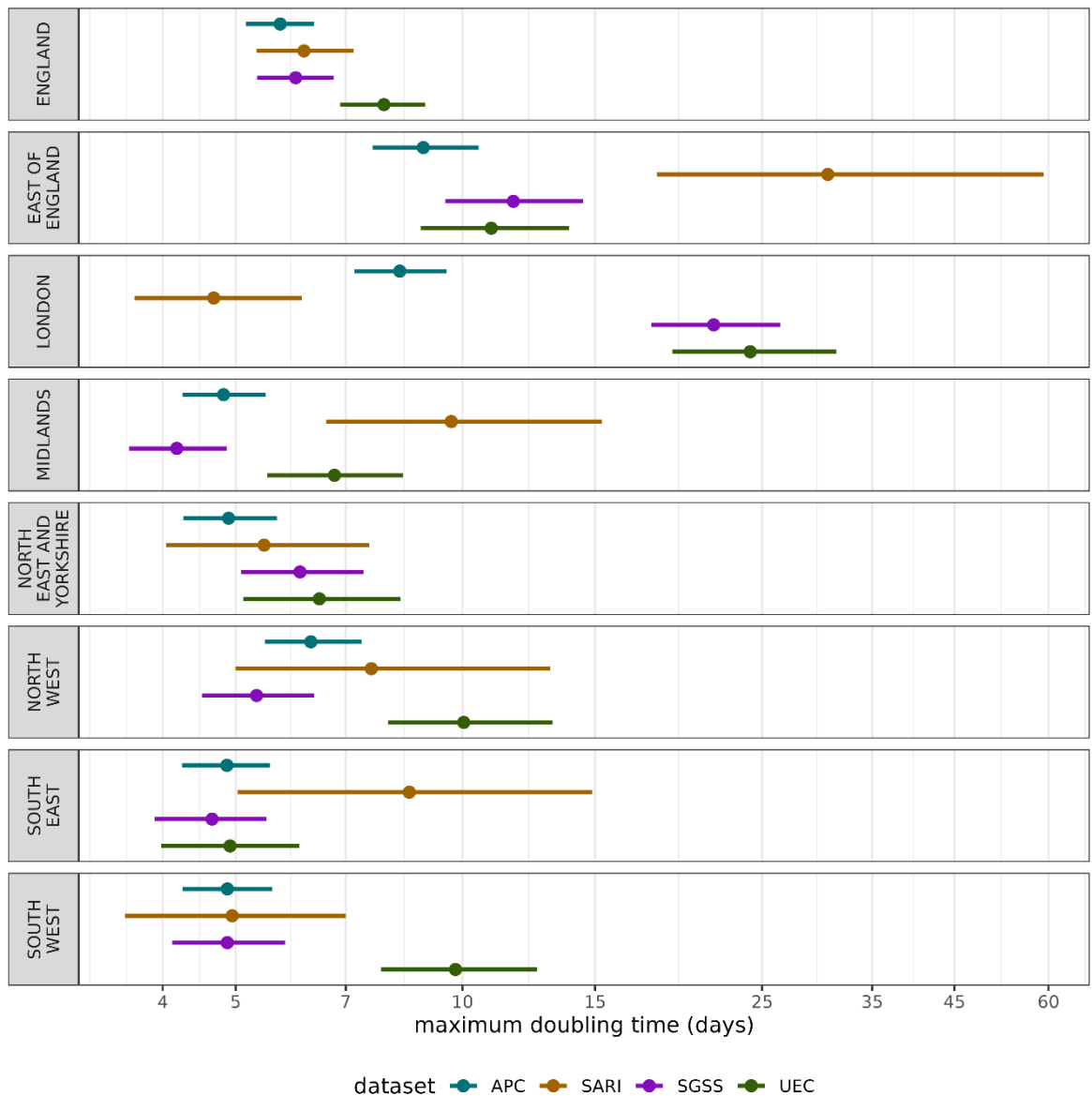

*Figure 7. The estimated doubling time (days until admissions double given the growth rate) across each data source over the 2022/23 winter season in England. The central point represents the median estimate, and the lines the 95% confidence interval. A lower doubling time corresponds to a faster growing epidemic.*

## Discussion

This research shows that while there may be variation in influenza admission rates across England, this variation can often be an artifact of the data studied, rather than just epidemiology. We have compared different datasets which use a range of definitions for an influenza hospital admissions and have different levels of population coverage, but there are many metrics and public health relevant questions where they agree. At a national level, for some of the analysis explored in this work there is strong agreement in trends, such as the maximum admission rate, estimated date of peak and maximum doubling time. However, this correspondence breaks down for the cumulative admissions rate, which may be more sensitive to the whole season surveillance.

The seasonal influenza wave in England in 2022/23 was an early wave, which spread quickly, peaked high compared to historic seasons and fell rapidly. Across datasets there is high variation in the timing of the influenza peak, though there is uncertainty in the modelled estimates. Across datasets there is the most regional variation in metrics for London and East of England. For example, in London the UEC and SGSS data sets are at very low levels, the sentinel SARI approach gives a large epidemic wave, whereas the retrospective but possibly more accurate APC gives a wave between the data sources. The result suggests that there are challenges in interpretation of the data in these regions.

Overestimates could be due to the sample containing disproportionate high reporting well engaged Trusts, whereas the lower estimates could be due to either lower ascertainment within the reporting hospitals, or issues with the reporting itself. This has real implications for monitoring the epidemic in real-time as APC is not available – the choice in dataset may lead to different conclusions in this region. It's crucial to note, except for APC, none of the datasets have near-complete coverage of secondary care providers in England. This is of course not expected of SARI which employs a sampling approach but does leave room for higher participation for Trusts within some regions in SARI. Consistently accurate values in more Trusts within the UEC would reduce the number of Trusts excluded and increase coverage, and in SGSS more Trusts providing organisation identifiers to would drive uncertainty down. Being able to link these data sources together, to associate a positive test with a patient's admission records would strengthen our understanding of ascertainment and the relationship between the collections. Where participation is high and data consistency is strong there is correspondence between the datasets increasing our confidence in the estimates derived. Notwithstanding London and the East of England regions, the SARI admission rates created using a fraction of the sample size of other datasets gives a robust national estimate, and when correcting for day-of-week effects a comparable regional estimate in many metrics. The higher cumulative season admission rates derived from SARI appear to be a result of higher non-peak-time values than other datasets, either representing better engagement/ascertainment in the participating Trusts, or a bias toward higher count returns.

For many regions, metrics and datasets results correspond highly, which is surprising given the different definitions for influenza admission, outlined in Table 1. While these definitions are different, their similar results imply that the reported estimates are robust and that influenza attributable admissions were well ascertained in the 2022 season. As some definitions strictly require test positivity, the inclusion of symptoms, and some clinical coding, this implies the testing for influenza is highly dependent on a symptom presentation and clinical coding dependent on testing, rather than clinical decision making alone. We would expect higher dependence on testing to define diagnosis given the confounding effect of other respiratory illnesses over the study period, namely COVID-19. The high correspondence in clinical coding and other definitions is interesting as coding itself can be so variable; the reproducibility of clinical coding between coders is limited [38] indicating local practices will differ between Trusts. This issue is exacerbated when comorbidities, which alters the risk for influenza patients [39] are taken into account, which causes more discrepancy in coding [40].

**Limitations**

To be able to compare data sources with different definitions of admission we have had to make assumptions about their relationships, such as inpatients testing positive being an admission, which is not strictly the case. Furthermore, to compare the datasets models were fit to the data to perform inference on the different quantities of interest and adjust for different reporting frequencies. This modelling step adds a source of bias into the results as the inferences are sensitive to how the models were fit. Ideally, this analysis would be conducted across several winter waves to understand changes in datasets over time, however, there was limited seasonal influenza incidence in 2020 and 2021 in England, and not all datasets existed before this time. Future research should consider a multi-year comparison of the influenza waves across the modern datasets available and relied upon in public health. The results presented are conducted using after-the-fact analysis to exclude specific poor reporting Trusts, which has implications for its utility in real-time. Without the inclusion criteria the results would be substantially more biased, as increased incorrect zero reporting would drag SGSS and UEC metrics downward. If reporting quality was perfect, with admissions reported representing true admissions across datasets we would expect that the difference in incidence would be an expression only of the difference in hospital admission definition and the Trusts included in reporting, however, this quality issue prevents us from inferring the differences in definitions. Crucially this work only looks at one very specific part of influenza surveillance. We have not explored the utility of different data sources from a public health perspective or an overall healthcare system view, but rather specifically on a comparable metric – the admission rate per capita. Further work should explore the wider data landscape for secondary care influenza in England more comprehensively in a qualitative manner.

**Conclusion**

In this research we have shown that there is correspondence in different influenza admission rate surveillance and data systems. This is shown by modelling the data to make each wave of incidence comparable in the 2022/23 season and estimating different key metrics of interest. Though there is clear agreement between the different data sources of varying sample size and collection rigor, this relationship breaks down regionally, particularly for London and East of England data reporting. The estimated peak size, timing and growth rates are similar across data sets, though the cumulative admission rate varies substantially regionally. We show that the choice of data can clearly impact the conclusions drawn from inference of the epidemic wave.

**Conflict of Interest**
The authors have declared that no competing interests exist. The authors were employed by the UKHSA but received no specific funding for this study.

**Data Availability Statement**
UKHSA operates a robust governance process for applying to access protected data that considers:
- the benefits and risks of how the data will be used
- compliance with policy, regulatory and ethical obligations

- data minimisation
- how the confidentiality, integrity, and availability will be maintained
- retention, archival, and disposal requirements
- best practice for protecting data, including the application of 'privacy by design and by default', emerging privacy conserving technologies and contractual controls

Access to protected data is always strictly controlled using legally binding data sharing contracts. UKHSA welcomes data applications from organisations looking to use protected data for public health purposes.

To request an application pack or discuss a request for UKHSA data you would like to submit, contact [DataAccess@ukhsa.gov.uk](DataAccess@ukhsa.gov.uk).

# References

[1] UKHSA, "National flu and COVID-19 surveillance reports: 2022 to 2023 season," 2022. [Online]. Available: https://www.gov.uk/government/publications/sources-of-surveillance-data-for-influenza-covid-19-and-other-respiratory-viruses/sources-of-surveillance-data-for-influenza-covid-19-and-other-respiratory-viruses. [Accessed 14 November 2022].

[2] NHS Digital, "Hospital Episode Statistics (HES)," 2022. [Online]. Available: https://digital.nhs.uk/data-and-information/data-tools-and-services/data-services/hospital-episode-statistics#:~:text=Hospital%20Episode%20Statistics%20(HES)%20is%20a%20database%20containing%20details%20of,Commissioning%20Data%20Set%20(CDS). [Accessed 15 November 2022].

[3] W. A. Fischer, M. Gong, S. Bhagwanjee and J. Sevransky, "Global burden of influenza as a cause of cardiopulmonary morbidity and mortality," *Global Heart,* vol. 9, no. 3, pp. 325-336, 2014.

[4] D. E. Morris, D. W. Cleary and S. C. Clarke, "Secondary bacterial infections associated with influenza pandemics," *Frontiers in microbiology,* vol. 8, 2017.

[5] C. E. Troeger, B. F. Blacker, I. A. Khalil, S. R. Zimsen, S. B. Albertson, D. Abate, J. Abdela, T. B. Adhikari, S. A. Aghayan and S. Agrawal, "Mortality, morbidity, and hospitalisations due to influenza lower respiratory tract infections, 2017: an analysis for the Global Burden of Disease Study 2017," *The Lancet Respiratory Medicine,* vol. 7, no. 1, pp. 69-89, 2019.

[6] D. Fleming, R. Taylor, F. Haguinet, C. Schuck-Paim, J. Logie, D. Webb, R. Lustig and G. Matias, "Influenza-attributable burden in United Kingdom primary care," *Epidemiology & Infection,* vol. 144, no. 3, pp. 537-547, 2016.

[7] J. R. Ortiz, K. M. Neuzil, C. R. Cooke, M. B. Neradilek, C. H. Goss and D. K. Shay, "Influenza pneumonia surveillance among hospitalized adults may underestimate the burden of severe influenza disease," *PloS one,* vol. 9, no. 11, 2014.

[8] WHO, "Influenza (Seasonal)," 2018. [Online]. Available: https://www.who.int/news-room/fact-sheets/detail/influenza-(seasonal). [Accessed 14 November 2022].

[9] NHS, "Flu," 2022. [Online]. Available: https://www.nhs.uk/conditions/flu/ . [Accessed 14 November 2022].

[10] Center for Disease Control, "U.S. Influenza Surveillance: Purpose and Methods," 2022. [Online]. Available: https://www.cdc.gov/flu/weekly/overview.htm#HHSProtect. [Accessed 16 December 2022].

[11] WHO, "Global Influenza Programme," 2022. [Online]. Available: https://www.who.int/teams/global-influenza-programme/surveillance-and-monitoring. [Accessed 2015 November 2022].

[12] ECDC, "Risk groups for severe influenza," 2022. [Online]. Available: https://www.ecdc.europa.eu/en/seasonal-influenza/prevention-and-control/vaccines/risk-groups. [Accessed 1 December 2022].

[13] Y. Qi, J. Shaman and S. Pei, "Quantifying the Impact of COVID-19 Nonpharmaceutical Interventions on Influenza Transmission in the United States," *The Journal of infectious diseases,* vol. 224, no. 9, pp. 1500-1508, 2021.

[14] Q. S. Huang, T. Wood, L. Jelley, T. Jennings, S. Jefferies, K. Daniells and A. Nesdale, "Impact of the COVID-19 nonpharmaceutical interventions on influenza and other respiratory viral infections in New Zealand," *Nature communications,* vol. 12, no. 1, pp. 1-7, 2021.

[15] H. Lei, M. Xu, X. Wang, Y. Xie, X. Du, T. Chen, L. Yang, D. Wang and Y. Shu, "Nonpharmaceutical interventions used to control COVID-19 reduced seasonal influenza transmission in China," *The Journal of infectious diseases,* vol. 222, no. 11, pp. 1780-1783, 2020.

[16] A. Gimma, J. D. Munday, K. L. M. Wong, P. Coletti, K. v. Zandvoort, K. Prem, C. C.-1. w. group, P. Klepac, G. J. Rubin, S. Funk, W. J. Edmunds and C. I. Jarvis, "Changes in social contacts in England during the COVID-19 pandemic between March 2020 and March 2021 as measured by the CoMix survey: A repeated cross-sectional study," *PLoS medicine,* vol. 19, no. 3, 2021.

[17] J. Nazareth, D. Pan, C. A. Martin, I. Barr, S. G. Sullivan, I. Stephenson, A. Sahota, T. W. Clark, L. B. Nellums, J. W. Tang and M. Pareek, "Is the UK prepared for seasonal influenza in 2022-23 and beyond?," *The Lancet Infectious Diseases,* vol. 22, no. 9, pp. 1280 - 1281, 2022.

[18] J. T. Kubale, A. M. Frutos, A. Balmaseda, S. Saborio, S. Ojeda and C. Barilla, "High co-circulation of influenza and SARS-CoV-2," *medRxiv : the preprint server for health sciences,* 2022.

[19] T. Chotpitayasunondh, T. K. Fischer, J.-M. Heraud and A. C. Hurt, "Influenza and COVID-19: What does co-existence mean?," *Influenza and other respiratory viruses,* vol. 15, no. 3, pp. 407-412, 2021.

[20] NHS Digital, "How we collect and process Hospital Episode Statistics (HES) data," 2018. [Online]. Available: https://digital.nhs.uk/data-and-information/data-tools-and-services/data-services/hospital-episode-statistics/how-we-collect-and-process-hospital-episode-statistics-hes-data. [Accessed 15 November 2022].

[21] A. Herbert, L. Wijlaars, A. Zylbersztejn, D. Cromwell and P. Hardelid, "Data resource profile: hospital episode statistics admitted patient care (HES APC)," *International journal of epidemiology,* vol. 46, no. 4, pp. 1093-1093, 2017.

[22] WHO, "ICD-10 Version:2010," 2010. [Online]. Available: https://icd.who.int/browse10/2010/en. [Accessed 15 November 2022].

[23] UKHSA, "Second Generation Surveillance Service," 2022. [Online]. Available: https://sgss.phe.org.uk/Security/Login. [Accessed 21 December 2022].

[24] NHS Digital, "The processing cycle and HES data quality," 2022. [Online]. Available: https://digital.nhs.uk/data-and-information/data-tools-and-services/data-services/hospital-episode-statistics/the-processing-cycle-and-hes-data-quality#sus-and-hes-data-quality. [Accessed 2022 November 17].

[25] National Institute for Health and Care Excellence, "How should I diagnose seasonal influenza?," NICE, January 2023. [Online]. Available: https://cks.nice.org.uk/topics/influenza-seasonal/diagnosis/diagnosis/.

[26] Public Health England, "Protocol for the Surveillance of Severe Acute Respiratory Infections in England (SARI-Watch) – 2020/21," 2020.

[27] NHS, "Urgent and Emergency Care Daily Situation Reports 2021-2022," 2022. [Online]. Available: https://www.england.nhs.uk/statistics/statistical-work-areas/uec-sitrep/urgent-and-emergency-care-daily-situation-reports-2021-22/ . [Accessed 23 November 2022].

[28] NHS, "Submit your daily situation report (sitrep)," 2022. [Online]. Available: https://www.england.nhs.uk/submit-your-daily-situation-report-sitrep/ . [Accessed 23 November 2022].

[29] NHS England, "Process and definitions for the daily situation report," 2021. [Online]. Available: https://www.england.nhs.uk/publication/process-and-definitions-for-the-daily-situation-report/. [Accessed 14 November 2022].

[30] UK Government, "The Health Protection (Notification) Regulations 2010," 2010. [Online]. Available: https://www.legislation.gov.uk/uksi/2010/659/schedule/2/made.

[31] UK Health Security Agency, "Laboratory reporting to UKHSA," May 2023. [Online]. Available: https://assets.publishing.service.gov.uk/government/uploads/system/uploads/attachment_data/file/1159953/UKHSA_Laboratory_reporting_guidelines_May_2023.pdf. [Accessed 01/10/2023].

[32] V. Dardanoni, M. Laudicella and P. L. Donni, "Hospital Choice in the NHS," *Health, Econometrics and Data Group (HEDG) Working Papers,* vol. 18, no. 4, 2018.

[33] S. Arora, R. C. Cheung, Sherlaw-Johnson, C. Hargreaves and D. S. Hargreaves, "Use of age-specific Hospital catchment populations to investigate geographical variation in inpatient admissions for children and young people in England: retrospective, cross-sectional study," *BMJ open,* vol. 8, no. 7, 2018.

[34] Office for Health Improvement and Disparities, "NHS Acute (Hospital) Trust Catchment Populations," Office for Health Improvement and Disparities, 2022. [Online]. Available: https://app.powerbi.com/view?r=eyJrIjoiODZmNGQ0YzItZDAwZi00MzFiLWE4NzAtMzVmNTUwMThmMTVlIiwidCI6ImVlNGUxNDk5LThmMzUtNGIyZS1hZDQ3LTVmM2NmOWRlODY2NiIsImMiOjh9.

[35] NHS Digital, "Estates Returns information Collection," NHS Digital, 13 October 2022. [Online]. Available: https://digital.nhs.uk/data-and-information/publications/statistical/estates-returns-information-collection. [Accessed 2023].

[36] S. N. Wood, "Fast stable restricted maximum likelihood and marginal likelihood estimation of semiparametric generalized linear models," *Journal of the Royal Statistical Society (B),* vol. 73, no. 1, pp. 3-36, 2011.

[37] G. L. Simpson, "gratia: Graceful ggplot-Based Graphics and Other Functions for GAMs," Feburary 2023. [Online]. Available: https://gavinsimpson.github.io/gratia/.

[38] J. Dixon, C. Sanderson, P. Elliott, P. Walls, J. Jones and M. Petticrew, "Assessment of the reproducibility of clinical coding in routinely collected hospital activitydata: a study in two hospitals," *Journal of Public Health,* vol. 20, no. 1, pp. 63-69, 1998.

[39] D. L. Schanzer and T. W. T. Joanne M Langley, "Co-morbidities associated with influenza-attributed mortality, 1994–2000, Canada," *Vaccine,* vol. 25, no. 36, pp. 4697-4703, 2008.

[40] Department of Health and Social Care, "Payment by Results in the NHS: data assurance," GOV.UK, 2013.